\documentclass{amsart}
\usepackage{amsfonts}
\usepackage{amsmath}
\usepackage[latin1]{inputenc}
\usepackage{amsfonts}

\newtheorem{theorem}{Theorem}
\newtheorem{lemma}{Lemma}
\newtheorem{corollary}{Corollary}
\newcommand{\lp}{\mathcal{LP}}

\begin{document}

\title[Lee-Yang measures and wave functions]
{Lee-Yang measures and wave functions}

\author{Dimitar K. Dimitrov}
\address{Departamento de Matem\'atica Aplicada, IBILCE, Universidade Es\-ta\-du\-al Paulista,
15054-000 S\~{a}o Jos\'{e} do Rio Preto, SP, Brazil}
\email{dimitrov@ibilce.unesp.br}

\thanks{Research supported by the Brazilian Science Foundations CNPq under Grant 307183/2013-0 and FAPESP under Grant 09/13832-9.}

\subjclass[2010]{42A38, 82C20, 33C45}

\keywords{Lee-Yang measure, Laplace transform, Fourier transform, Wronskian, Slater determinant, wave function}

\begin{abstract}
We establish necessary and sufficient conditions for a Borel measure to be a Lee-Yang one which means that 
its Fourier transform possesses only real zeros. Equivalently, we answer a question of P\'olya 
who asked for a characterisation of those positive positive, even and sufficiently fast decaying kernels whose 
Fourier transforms have only real zeros. The characterisation is given in terms of Wronskians of polynomials that 
are orthogonal with respect to the measure. The results show that Fourier transforms of a rather general class 
of measures can be approximated by symmetrized  Slater determinants composed by orthogonal 
polynomials, that is, by some wave functions which are symmetric like the Boson ones. Brief comments on 
possible interpretation and applications of the main results in quantum and statistical mechanics, to 
Toda lattices and the general solution of the heat equation, are given.  We discuss briefly the possibility 
of represent the Riemann $\xi$ function as a partition function of a statistical mechanics system.     
\end{abstract}

\maketitle

\section{Introduction}

Suppose that $K$ is a positive kernel which decays sufficiently fast at $\pm \infty$, say it is in the 
Scwartz class, and its Fourier transform 
$$
\mathcal{F}(z;K) = \int_{-\infty}^{\infty} e^{-izt} K(t) dt
$$
is an entire function. More generaly, we consider positive Borel measures $d\mu$ with the property that 
$$
\mathcal{F}_\mu(z) = \int_{-\infty}^{\infty} e^{-izt}  d\mu(t)
$$
defines an entire function. The problem to characterise the measures $\mu$ for which $\mathcal{F}_\mu$ 
has only real zeros has been of interest both in mathematics, because of the Riemann hypothesis, and in 
physics, because of the validity of the so-called general Lee-Yang theorem for such measures.

Is seems that P\'olya was the first to formulate the problem explicitly in his very short note \cite{Pol26} 
which begins with the following sentences (we only change the notation for the functions): ``What properties 
of the function $K(u)$ are sufficient to secure that the integral 
$$
2 \int_{-\infty}^{\infty} K(u) \cos(zu) du = \mathcal{F}(z)
$$ 
has only real zeros? The origin of this rather artificial question is the Riemann hypothesis concerning the 
Zeta-function. If we put
$$
K(u) = \Phi(u) := \sum_{n=1}^{\infty} (4 \pi^2 n^4 e^{9u/4} - 6 \pi n^2 e^{5u/4}) e^{-n^2 \pi e^{2u}}
$$
$\mathcal{F}(z)$ becomes Riemann's function $\xi(z)$.'' P\'olya adopted Riemann's original definition 
of $\xi$ in terms of the zeta function,
$$
\xi(t) = \Gamma(s/2) (s-1) \pi^{-s/2} \zeta(s),\ \ s=1/2+it.
$$
Since $\Phi$ is an even kernel which decreases extremely fast, the above definition of P\'olya for $\mathcal{F}$, 
in the case when $K$ is even, is exactly the one for the Fourier transform.    
The Riemann's hypothesis, as formulated by himself, states that the zeros of $\xi$ are all real.  The efforts to 
settle the Riemann hypothesis establishing the fact that $\xi$, define as Fourier transform has failed despite 
the efforts of P\'olya \cite{PolC}, de Bruijn \cite{Bru50} and many other celebrated mathematician for two reasons. The first one is that  
the above question of P\'olya still remains open. The second is that the sufficient conditions for the kernels that 
have been proved either are extremely difficult to be for verified for $\Phi$  or simply do not hold for it.       

Aiming to understand the phase transition in the classical Ising model, Lee and Yang established a beautiful 
result shown that the zeros of some polynomials lie on the imaginary axis. Consider the set $\Lambda \subset 
\mathbb{R}^d$ of $N$ lattice sites. With every site $j\in \Lambda$ we associate a magnetic spin and a spin 
variable $\sigma_j \in {-1,1}$. Then any vector $\mathbf{\sigma}=(\sigma_1,\ldots,\sigma_N)$ describes 
a spin configuration. In the Ising model the nearest neighbour spins and only they interact. If $j,k \in \Lambda$ 
are such neighbours the interaction is denoted by $J_{jk}$ and the model is said to be ferromagnetic if all 
interactions $J_{jk}$ are positive. Moreover, for each $j$, 
an external magnetic field $h_j$ acts on the spin at that site. The Hamiltonian of the Ising model described 
in this manner is given by
$$
H(\sigma) = - \sum_{n n } J_{jk} \sigma_j \sigma_k - \sum_{j \in \Lambda} h_j \sigma_j,
$$ 
where the first sum is extended over the nearest neighbour sites.  Then the corresponding partition function 
is given by 
$$
Z_\beta (h_1,\ldots, h_N) = \sum_{\sigma} e^{-\beta H(\sigma)},
$$
where $\beta$ is the inverse temperature and the sum is over all spin configurations $\sigma$. It is clear 
then that the number of terms is $2^N$. Lee and Yang \cite{LY, YL} proved for any $\beta>0$ the above
partition function of the ferromagnetic Ising model obeys the property that 
\begin{equation}
\label{LYP}
 Z_\beta (h_1,\ldots, h_N)\ \ \ \mathrm{when\ all}\ \ h_j \in \mathbb{H}_+ = \{ h\in \mathbb{C}\ :\ \Re\, h >0\}
\end{equation}
and, in particular, all zeros of $Z_\beta (h,\ldots, h)$ belong to the imaginary axis. This is the celebrated 
Lee-Yang theorem, also called the circle theorem because the straightforward change of variable $z=e^{\beta h}$
transforms the partition function onto $\tilde{Z}(z)=z^{-N} \tilde{P}_{2n}(z)$, where $\tilde{P}_{2N}$ is an algebraic 
polynomial whose zeros lie on the unit circle $|z|=1$. The theorem of Lee and Yang has been 
extended and generalised in various directions, starting from Asano \cite{Asa68}, Ruelle \cite{Rue71}, Simon and 
Griffiths \cite{SimGri73}, Newman \cite{New74}, Lieb and Sokal \cite{LieSok81}, up to the very recent results of Fr\"{o}hlich and 
Rodriguez \cite{FroRod12}  and of Lebowitz, Ruelle and E. R. Speer \cite{LebRueSpe12}, just to mention a few.

It is clear that the partition function $Z_\beta$ can be written in the form 
\begin{equation}
\label{ZInt}
Z_\beta (h_1,\ldots, h_N) = \int e^{-\beta H} d\mu_1(\sigma_1)\ldots  d\mu_N(\sigma_N),
\end{equation} 
with 
$$
d\mu(x) = \frac{1}{2} \left( \delta(x+1) + \delta(x-1) \right),\ \ \ j=1,\ldots, N.
$$
Newman \cite{New74} suggested to consider the partition function for much wider class of measures $d\mu$ obeying the following 
properties:
\begin{itemize}
\item[(i)] $d\mu$ is an even Borel measure on $\mathbb{R}$;
\item[(ii)] $\int \exp(bx^2) d\mu(x) < \infty$ for every $b\in \mathbb{R}$;
\item[(iii)] $\int \exp(hx) d\mu(x) \neq 0$ for every $h \in \mathbb{H}_+$.
\end{itemize}
Then Newmann proved that once all $d\mu_j$ satify these requirements, the partition function  defined by (\ref{ZInt})
obeys the property (\ref{LYP}). A measure $d\mu$ is called a Lee-Yang measure if it satisfies (i), (ii) and (iii). Obviously (iii) 
is equivalent to the fact that all zeros of the formal Laplace transform 
$$
\mathcal{L}_\mu(z) = \int e^{-zt} d\mu(t)
$$
lie on the imaginary axis.  We shall use this formalism to avoid writing the Fourier transform with imaginary arguments and 
because this is the general form of a partition function considered as a function of the fugacity variable $z$. 
With the foregoing definition Newman's result states that the Lee-Yang properties (\ref{LYP}) 
hold for the partition function (\ref{ZInt}) of a ferromagnetic Ising model provided all measure measures $d\mu_j$ are 
Lee-Yang ones.  This raises the natural question for characterising the Lee-Yang measures. 

The identity $\mathcal{L}_\mu(iz) = \mathcal{F}_\mu(z)$ shows that the latter problem is equivalent to P\'olya's 
one. We provide necessary and sufficient conditions for a measure to be a Lee-Yang measure for a much wider class of 
Borel measures $d\mu$ than described by the requirement (ii). The only essential restriction we impose is that 
the moments of the measure do not increase very fast so that the Fourier transform $\mathcal{F}_\mu$ is an entire function and that the 
moment functional is a positive definite one, so that the sequence of polynomials $\{p_n\}$, orthonormal with respect to the 
measure,
$$
\int p_j(x) p_n(x) d\mu(x) = \delta_{jn},\ \ \ \ deg(p_n) =n,
$$
exist and are uniquely determined. Then our main result states that the measure $d\mu$ is a Lee-yang measure if and only if the 
all zeros of all Wronskians $W(p_1,\ldots, p_n;z)$ are purely imaginary.  In fact these Wronskian are the correct 
polynomial approximations of the Fourier (of the formal Laplace) transform which ``recognise" the location of the zeros 
of these  transforms because, as we shall see, they are related to the so called Jensen polynomials of $\mathcal{L}_\mu(iz)$
and $\mathcal{F}_\mu(z)$. While the univariate version of the main result reveals this unexpected relation only, the natural procedure of multivariate 
extension, which was employed by Lee and Yang themselves, yields the very nice multivariate polynomials (\ref{WW}). They are 
quotients of Slater determinants of the first $n$ orthogonal polynomials, starting with not with $p_0$ but with $p_1$, with 
variables $z_1,\ldots, z_n$ and the Vandermonde determinant of these variables. Thus it turns out to be a very interesting 
symmetrized wave function (SWF) which, from one side
is very similar to the fermion wave function because contains Slater determinants but on the other one, it is symmetric, so that 
it resembles a boson multiparticle wave function. However the boson wave functions are simply permanents while here the 
symmetrization is different.  Then our result implies a rather curious phenomena: the formal Laplace transform of a measure, which is 
a partition function of a general Ising model, in the sense of Newman \cite{New74} can be obtained from the SWF 
by two consecutive limit processes. The first one one is a ``coalescence" of the ``position variables"  and the second is simply the 
thermodynamic limit, where to let $n$ to converge to infinity.

In the next section we provide the necessary definitions and formulate the main theorems. Section 3 contains preliminary results 
and the proofs of the statements. In Section 4 we comment possible interpretations and applications and emphasise 
on the interplay between the Riemann $\xi$ function and the Lee-Yang theorem.   

\section{Statements of the main results}

We begin with the necessary definitions related to the moment problem and orthogonal polynomials 
following the classical books of Shohat and Tamarkin \cite{ShoTam} and Szeg\H{o} \cite{Sze75}.  
Let $\{m_k\}_{k=0}^\infty$ be a sequence of real numbers  with $m_{2k}>0$ and $m_{2k+1}=0$, $k=0,1\ldots$. The $(n+1)\times(n+1)$ Hankel determinant 
$D_n$ is the one whose entries $d_{j\ell},\ j, \ell =0,\ldots,n$ are $d_{j\ell}=m_{j+\ell}$. If 
\begin{equation}
\label{Dn}
D_n > 0\ \  \mathrm{for\ every}\ \ n\in \mathbb{N} \cup \{0\},
\end{equation}
then there is an even Borel measure $d\mu$ whose support is an infinite subset of $\mathbb{R}$ and 
\begin{equation}
\label{MP}
m_j = \int_{-\infty}^{\infty} t^j d\mu(t).
\end{equation}
In other words, the Hamburger moment problem is solvable.
We consider even moment sequences $\{m_k\}$ which satisfy this requirement and the additional one: 
\begin{equation}
\label{Emu}
\limsup_{n \rightarrow \infty}  \frac{\mu_{2n}^{1/(2n)}}{2n} = 0.
\end{equation}
Since the formal series expansion of $\mathcal{L}_\mu(z)$ is 
$$
\sum_{k=0}^{\infty} \frac{m_{2k}}{(2k)!}\, z^{2k},
$$
then (\ref{Emu}) guarantees that it represents an entire function. Equivalently, 
$$
\mathcal{F}_\mu(z) = 
\sum_{k=0}^{\infty} (-1)^k \frac{m_{2k}}{(2k)!} z^{2k},
$$
is entire.
Moreover, results 
of Carleman, Hamburger and Stridsberg (see \cite[Theorem 1.11]{ShoTam}) state that the Hamburger 
moment problem is determined if 
$$
\limsup_{n \rightarrow \infty}  \frac{\mu_{2n}^{1/(2n)}}{2n} <  \infty.
$$
Therefore, 
if  (\ref{Emu}) holds, then the measure is uniquely determined and the orthonormal polynomials 
are represented by 
\begin{equation}
\label{OP}
p_n(x) =   
 (D_{n-1} D_{n})^{-1/2}  D_n(x),\ \  with\ \ D_n(x) = 
\left|   \begin{array}{cccccc}
m_0 & m_1 & \cdots & \cdots & m_n \\
m_1 & m_2 & \cdots & \cdots & m_{n+1} \\
\vdots & \vdots & \vdots & \vdots & \vdots \\
m_{n-1} & m_n & \vdots & \vdots & m_{2n-1} \\
1& x & \cdots & \cdots & x^n \\
\end{array} \right|.
\end{equation}

The Wronskian of the sufficiently smooth functions $f_1,\ldots, f_n$ is
$$
W(f_1,\ldots, f_n; z) = \left|   \begin{array}{ccccc}
f_1(z) & f_2(z) & \cdots & f_n(z) \\
f_1^\prime(z) & f_2^\prime(z) & \cdots  & f_n^\prime(z) \\
\vdots & \vdots & \vdots &  \vdots \\
f_1^{(n-1)}(z) & f_2^{(n-1)}(z) & \cdots & f_n^{(n-1)}(z) \\
\end{array} \right|.
$$
The Wrosnkians 
\begin{equation}
\label{W1}
W(p_1,\ldots, p_n; z),
\end{equation}
and $W(p_2,\ldots, p_n; z)$, composed by the orthogonal polynomials starting from the one of either the first or second  
degree play fundamental role in our results.

For every $n \in \mathbb{N}$ we define $C_n = \prod_{k=1}^{n-1} k!$. The 
constant $C_n$ can be expressed in terms of Barnes' double Gamma function 
$G$, $C_n=G(n+1)$. 

\begin{theorem} 
Let  $\{m_k\}_0^\infty$ be a moment sequence with the properties 
(\ref{Dn}) and (\ref{Emu}). Then the unique Borel measure $d\mu$ which is the 
solution of the moment problem (\ref{MP}) is such that its Fourier transform 
$\mathcal{F}_\mu$ possesses only real zeros (equivalently, all the zeros of its formal Laplace transform 
$\mathcal{L}_\mu$ are on the imaginary axis) if and only if, for every $n\in \mathbb{N}$,
the zeros of the Wronskian $W(p_1,\ldots,p_n; z)$ lie on the imaginary axis. Moreover,
$$
\frac{\sqrt{m_0 D_{2n}}}{C_{2n}}\, \left(\frac{z}{2n} \right)^{2n} \, W\left(p_1,\ldots, p_{2n};  \frac{2n}{z}\right) \rightarrow \mathcal{L}_\mu(z),
$$
$$
\frac{\sqrt{m_0 D_{2n}}}{C_{2n}}\, \left(\frac{iz}{2n}\right)^{2n}\, W\left(p_1,\ldots, p_{2n}; - \frac{2in}{z}\right) \rightarrow \mathcal{F}_\mu(z),
$$
and the convergence is uniform on the compacts subsets of the complex plane.
\end{theorem} 

Before we proceed with the next result, observe that the above Wronskians $W(p_1,\ldots,p_n; z)$ are polynomial 
of degree $n$ and the polynomials which converge to the Fourier transforms are the ``reversed" of $W(p_1,\ldots,p_n; z)$. 
It will become clear that the leading coefficient of $W(p_1,\ldots,p_n; z)$ is exactly $(-1)^n m_n$. Since the odd moments 
$m_{2m+1}$ are equal to zero, the polynomials reversed to $W(p_1,\ldots,p_{2m+1}; z)$ are not well defined, at least in the present 
setting. However, $m_{2n}>0$ because the measure is positive. Then the leading coefficients of the even degree Wronskians 
$W(p_1,\ldots,p_{2n}; z)$ are all positive and their reversed polynomials are well defined. In fact, the result can be rewritten in 
a straightforward manner performing the quadratic transformation $t^2=t_1$ in the definition of the Fourier transform $\mathcal{F}_\mu$
(equivalently, $u^2=u_1$ in P\'olya's setting) considering the Stieltjes moment problem for a measure on the positive half-line and 
dealing with the corresponding orthogonal polynomials. However, we prefer the approach with even moment sequences and 
measures because of the fact that both in the Riemann $\xi$ function and in Newman's setting about the general Lee-Yang 
theorem the measures are even.

Another necessary and sufficient condition is given in terms of the Wronskians $W(p_2,\ldots, p_n;z)$ (the first orthogonal polynomial which appears in the Wronskian is the orthonormal polynomial of degree two):

\begin{theorem} 
Let  $\{m_k\}_0^\infty$ be an even moment sequence which satisfies (\ref{Dn}) and (\ref{Emu}). Then the 
Fourier transform of  the measure $d\mu$ has only real zeros if and only if 
$$
W(p_2,\ldots, p_n;ix) <0 \ \  \ \mathrm{for\ every}\ \ x \in \mathbb{R}\setminus \{0\}\ \ \mathrm{and\ for\ each}\ \ n \geq 2.
$$
\end{theorem} 

As it will become clear, the latter Wrosnkains and those which appear in Theorem 1 are related by 
$$
W(p_2,\ldots, p_n;x) = - a_n W^2(p_1,\ldots, p_{n+2};x) + b_n W(p_1,\ldots, p_{n+1};x)\, W(p_1,\ldots, p_{n+3};x), 
$$
where $a_n$ and $b_n$ are positive and depend only on $n$ and on the moments $m_k$. 

Now we define the symmetrized Slater determinants of $p_1,\ldots, p_n$ by 
\begin{equation}
\label{WW}
\mathcal{W}_n (\mathbf{z}) = \mathcal{W}_n(z_1, z_2, \cdots, z_n) =
\frac{
\left|   \begin{array}{cccc}
p_1(z_1)& p_2(z_1) & \cdots  & p_n(z_1) \\
p_1(z_2)& p_2(z_2) & \cdots  & p_n(z_2) \\
\vdots & \vdots & \vdots & \vdots  \\
p_1(z_n)& p_2(z_n) & \cdots & p_n(z_n) \\
\end{array} \right|}
{
\left|   \begin{array}{cccc}
1 & z_1 & \cdots  & z_1^{n-1} \\
1 & z_2 & \cdots  & z_2^{n-1} \\
\vdots & \vdots & \vdots  & \vdots \\
1& z_n & \cdots &   z_n^{n-1} \\
\end{array} \right|
}\, .
\end{equation}
First of all, it is obvious that $\mathcal{W}_n(z,z,\ldots,z)= W(p_1,\ldots,p_{n}; z)$
Let us write $\mathcal{W}_n$ in the succinct form 
$$
\mathcal{W}_n (\mathbf{z}) =\frac{\mathcal{S}_n (p_1,\ldots,p_n; \mathbf{z})}{V_n(\mathbf{z})},
$$
where $\mathcal{S}_n$ is the Slater determinant in the numerator and $V_n(\mathbf{z})$ is the Vandermonde determinant. 
It is clear that $\mathcal{W}_n (\mathbf{z})$ is a multivariate polynomial because $\mathcal{S}_n$ vanishes whenever 
$z_j=z_k$ and then contains all factors $z_j-z_k$, $j\neq k$, and $V_n(\mathbf{z})$ is precisely the product 
of these factors. Consider both $\mathcal{S}_n$ and $V_n$ as polynomials of a specific variable, say $z_j$. Then obviously 
$deg(\mathcal{S}_n(z_j)) = n$ and $deg(V_n(z_j))=n-1$ which implies that $\mathcal{W}_n (\mathbf{z})$ is a multiaffine polynomial. 
On the other hand, every univariate polynomial can be uniquely extended to a multivariate affine polynomial. Hence 
$\mathcal{W}_n (\mathbf{z})$ is exactly the unique multiaffine extension of $W(p_1,\ldots,p_n; z)$. 

\begin{theorem} 
Let  $\{m_k\}_0^\infty$ be an even moment sequence with the properties (\ref{Dn}) and (\ref{Emu}) and 
$d\mu$ which is the corresponding Borel measure. Then the Fourier transform of $d\mu$ has only real zeros
if and only if every $\mathcal{W}_n (\mathbf{z})$, $n\in \mathbb{N}$, obeys the Lee-Yang property, i.e., 
$\mathcal{W}_n (\mathbf{z}) \neq 0$ if $z_j \in \mathbb{H}_+$, $j=1,\ldots, n$.
\end{theorem}

As we have already mentioned In the classical Ising model with $N$ spins the partition function naturally reduces to 
an algebraic polynomial. The reason is that the measure is a discrete one with only two mass points. 
The last statement reveal the importance of the wave polynomials $\mathcal{W}_n$ as 
natural approximations of the the partition function of a general Ising model where the measures $d\mu_j$ 
are very general positive Borel measures. 

Observe, however, that the result in Theorem 2 may be stated in terms of multivariate stable polynomials which have been 
studied extensively because of they fundamental role in various areas of mathematics. Recall that a multivariate polynomial 
$\mathcal{P}$ is stable if $\mathcal{P}(\mathbf{z}) \neq 0$ for every $\mathbf{z}$ with $\Im z_k>0$, or $k=1,\ldots,n$. We 
omit the formal equivalent statement because it is straightforward.

\section{Proofs}
First we provide information on various result we need for the proof of the main results.
\subsection{Functions in the Laguerre-P\'olya class and their Jensen and Appell polynomials}

The real entire function $\psi(x)$ belong to the
Laguerre-P\'olya class $\lp$ if it can be represented as
$$
\psi(x) = c x^{m} e^{-a x^{2} + b x}
\prod_{k=1}^{\infty} (1+x/x_{k}) e^{-x/x_{k}},
$$
where $c, b$ and $x_{k}$ are real, $a\geq 0$, $m\in \mathbb{N} \cup \{0\}$ and $\sum x_{k}^{-2} < \infty$. 
The functions in $\lp$ and only they obey the property that that are local uniform limits, that is, uniform limits
in the the compact subset of $\mathbb{C}$, of polynomials with only real zeros. Such polynomials are usually called 
hyperbolic ones. If the Maclaurin expansion 
of $\psi \in \lp$ is
\begin{equation}
\psi(x) = \sum_{k=0}^{\infty} \gamma_{k}
\frac{x^{k}}{k!}
\label{Maclaurin}
\end{equation} 
then its Jensen polynomials are defined by
$$
g_{n}(x) = g_{n}(\psi;x) := \sum_{j=0}^{n} {n\choose
j} \gamma_{j}
x^{j}, \ \ \  n = 0, 1, \ldots.
$$
Jensen considered also the polynomials 
$$
A_n(x) = A_n(\psi;x) = x^n g_{n}(1/x) = \sum_{j=0}^{n} {n\choose j}
\gamma_{j} x^{n-j}
$$
which are nowadays called the Appell polynomials of $\psi$. The generating functions of polynomials $g_n$ and $A_n$ 
are given in terms of $\psi$ (see \cite{CraCso}),
$$
e^z \psi(zt) = \sum_{n=0}^{\infty} g_n(t) \frac{z^n}{n!},
$$
$$
e^{zt} \psi(t) = \sum_{n=0}^{\infty} A_n(z) \frac{t^n}{n!}.
$$

Jensen himself proved the following fundamental 
characterisation of the functions in the Laguerre-P\'olya class:
\begin{lemma} 
A function $\psi$ with the Maclaurin expansion (\ref{Maclaurin}) belongs to $\lp$ if and only if all its Jensen  
polynomials $g_{n}(\psi;x)$, $n\in \mathbb{N}$ are hyperbolic. 
Moreover, the sequence  $\{ g_{n}(\psi;x/n) \}$ converges locally uniformly to
$\psi(x)$.
\end{lemma} 
Jensen's result implies immediately that the entire function (\ref{Maclaurin}), with $\psi(0)\neq 0$, is in 
$\lp$ if and only if all its Appell polynomials 
$A_n(\psi;x)$ are hyperbolic and that 
\begin{equation}
(x/n)^n A_n(\psi;n/x)  \rightarrow \psi(x)\ \ \mathrm{unformly\ on\ the\ compact\ subsets\ of}\ \ \mathbb{C}.
\label{AConv}
\end{equation} 
The beauty of these statements is that the among the various polynomial sequences that converge locally uniformly to $\psi$ 
the Jensen and the Appell polynomials are the best ones because they ``recognise" if $\psi$ is in the Laguerre-P\'olya class or not.

Another relevant result concerns the so-called Tur\'an determinants of the Jensen polynomials, defined by 
$$
\Delta_n(\psi;x) = g_{n}^2(\psi;x) - g_{n-1}^2(\psi;x) g_{n+1}^2(\psi;x).
$$
Craven and Csordas \cite{CraCso} (see also \cite{CsoVar}) proved the following:
\begin{lemma} 
Let the Maclaurin coefficients of the real entire function $\psi$ be such that $\gamma_{k-1}\gamma_{k+1}<0$ whenever $\gamma_k=0$, $k=1,2,\ldots$.
Then $\psi \in \lp$ if and only if 
\begin{equation}
\label{Ti}
\Delta_n(\psi ; x) > 0\ \  \ \mathrm{for\ every}\ \ x \in \mathbb{R}\setminus \{0\}\ \ \mathrm{and\ for\ each}\ \ n \in \mathbb{N}.
\end{equation}
\end{lemma}
The condition on the Maclaurin coefficients is essential in this statement but is hold in our specific application. 

\subsection{Wronskains of orthogonal polynomials} 
In 1960 Karlin and  Szeg\H{o} wrote an incredibly long paper where they treated questions concerning positivity of 
Wronskians and the so-called generalised Tur\'an determinants (or Turanians) of the classical families of orthogonal polynomials. 
The clue idea was to transform the Tutanians of one family to the Wronskians of another one and then study the positivity. 
More recently Leclerc \cite{Lec}  established a general expression for Wronskians of orthogonal polynomials. Since it as a clue 
relation which reveals the interlays between Wronskians and Jensen and Appell polynomials, we formulate Leclerc's result 
adapting it to our notations. For a moment sequence $\{m_k\}_0^\infty$, define the polynomials
$$
q_n(z) = \sum_{j=0}^{n} {n\choose j} m_j  (-z)^{n-j},\ \ \ n=0,1,\ldots.
$$
Leclerc's relation reads as follows:
\begin{lemma} 
The following identity holds for all integer values of $\ell, n \in \mathbb{N}$:
$$
W(D_\ell(z), D_{\ell+1}(z), \ldots, D_{\ell+n-1}(z))
= c_{\ell,n}
\left|   \begin{array}{ccccc}
q_n(z)& q_{n+1}(z) & \cdots  & q_{n+\ell-1}(z) \\
q_{n+1}(z)& q_{n+2}(z) & \cdots  & q_{n+\ell}(z) \\
\vdots & \vdots & \vdots & \vdots  \\
q_{n+\ell-1}(z)& q_{n+\ell}(z) & \cdots  & q_{n+2\ell-2}(z) \\
\end{array} \right|,
$$
where 
$$
c_{\ell,n} = (-1)^{n\ell}\, C_n \prod_{k=1}^{n-1} D_{k+\ell-1}.
$$
\end{lemma}
We formulate an important immediate consequence, in the case when $\ell=1$:
\begin{corollary} 
For every $n\in \mathbb{N}$ we have
\begin{equation}
(-1)^n\, q_n(z) = \frac{\sqrt{m_0 D_{n}}}{C_{n}}\, W(p_1,\ldots, p_n; z).
\end{equation}
\end{corollary}

\subsection{Proof of the main results}
Now the proofs of the main result are simply a combination of the facts we have exposed.
Since that the Maclaurin expansion of $\mathcal{L}_\mu$ is 
$$
\mathcal{L}_\mu(z) = \sum_{k=0}^\infty \frac{m_k}{k!} (-z)^k,
$$
then its Jensen and Appel polynomials are given by 
$$
g_n(\mathcal{L}_\mu;z) = \sum_{k=0}^n {n \choose k}\,  {m_k}\  (-z)^k 
$$
and 
\begin{equation}
\label{An}
A_n(\mathcal{L}_\mu;z) = \sum_{k=0}^n {n \choose k}\,  {m_k}\  (-z)^{n-k}. 
\end{equation}
Then 
$$
g_n(\mathcal{F}_\mu;z) = \sum_{k=0}^n {n \choose k} \, {m_k}\  (-iz)^k 
$$
and 
$$
A_n(\mathcal{F}_\mu;z) = \sum_{k=0}^n {n \choose k} \, {m_k}\  (-iz)^{n-k}.
$$
On the other hand, obviously
$$
A_n(\mathcal{L}_\mu;z) = (-1)^n q_n(z),
$$ 
and then by the last corollary 
\begin{equation}
\label{AnW}
A_n(\mathcal{L}_\mu;z) = \frac{\sqrt{m_0 D_{n}}}{C_{n}}\, W(p_1,\ldots, p_n; z).
\end{equation}
This yields 
$$
g_n(\mathcal{L}_\mu;z) = \frac{\sqrt{m_0 D_{n}}}{C_{n}}\, z^n\, W(p_1,\ldots, p_n;1/z)
$$
and the similar identities 
$$
A_n(\mathcal{F}_\mu;z) = \frac{\sqrt{m_0 D_{n}}}{C_{n}}\, W(p_1,\ldots, p_n; iz)
$$
and
$$
g_n(\mathcal{F}_\mu;z) = \frac{\sqrt{m_0 D_{n}}}{C_{n}}\, (iz)^n\, W(p_1,\ldots, p_n;-i/z),
$$
Now Theorem 1 follows from Lemma 1. 

We furnish the argument to prove Theorem 2. Leclerc's relation between Wronskians of orthogonal 
polynomials and the Appell polynomials, for $\ell=2$ shows that 
$$
W(p_2,\ldots,p_n;z) = - \tilde{c}_{n} \left( A_{n+2}^2(\mathcal{L}_\mu;z) - A_{n+1}(\mathcal{L}_\mu;z) A_{n+3}(\mathcal{L}_\mu;z) \right),
$$
where $\tilde{c}_{n}$ is portive and depends only on $n$ and on the moments $m_k$. Then 
\begin{eqnarray*}
W(p_2,\ldots,p_n;iz) = - \tilde{c}_{n} \left( A_{n+2}^2(\mathcal{F}_\mu;z) - A_{n+1}(\mathcal{F}_\mu;z) A_{n+3}(\mathcal{F}_\mu;z) \right),
\end{eqnarray*}
or equivalently, 
\begin{eqnarray*}
z^{2n+2}\, W(p_2,\ldots,p_n;i/z) = - \tilde{c}_{n} \left( g_{n+2}^2(\mathcal{F}_\mu;z) - g_{n+1}(\mathcal{F}_\mu;z) A_{n+3}(\mathcal{F}_\mu;z) \right).
\end{eqnarray*}
The requirement on the sequence $\{ \gamma_k \}$ in Lemma 2 is satisfied for $\psi=\mathcal{F}_\mu$, with even even measure because in this case 
$\gamma_{2k+1}=0$ and $\gamma_{2k}=(-1)^k\, m_{2k}\, k!/(2k)!$.  

The statement of Theorem 3 follows immediately from Theorem 1 by the general argument that the 
multivariate affine extension $\mathcal{R}_n$ of a univariate polynomial $r_n$ obey the Lee-Yang property 
if and ably if the zeros of $r_n$ are purely imaginary. Lee-Yang property in this setting is 
$$
\mathcal{W}_n(\mathbf{z}) \neq 0\ \ \mathrm{when}\ \  z_j \in \mathbb{H}_+.
$$
In fact, this argument holds for any circular region and relies on a theorem of Grace and the Walsh 
incidence theorem. In particular, we may consider the upper half plane and stable polynomials.  Lee and Yang 
themselves employed the ingenious idea of extending univariate polynomials to multivariate ones. In their case 
the region was the unit disc so that they considered polynomials which are commonly know as Schur stable 
polynomials. In fact, it seems that Lee and Yang were first to use the idea so successfully so they might be considered
the scientists who have developed it.    

We skip the further details and refer to the paper of Ruelle \cite{Rue10} and Borcea and Br\"and\'en \cite{BorBra1, BorBra2}
for more information about multivariate stable polynomials. 

\section{Possible interpretations and applications} 

We begin this section with the obvious comment that our results provide necessary and sufficient conditions 
for the Riemann hypothesis. We avoid stating it explicitly for the reason that it would be another 
equivalent form of Riemann's celebrated conjecture. The interplay between the Riemann $\xi$ function 
and the Lee-Yang has been discussed widely and we add a short comments at the end of this paper.

First we sketch some natural questions and ideas which arise from the surprising explicit form of the Jensen 
polynomials as Wrosnkians of orthogonal polynomials, the corresponding ``lifting" of the Wronskian to the ``wave function" 
$\mathcal{W}_n$ and the limit process. The idea of approximation the Fourier transform by polynomials which converge 
locally uniformly to it is itself surprising and a bit odd from a first glance but one is led naturally to it by the results in Theorem 1. 

\subsection{The wave function} 
It is well known that the classical orthogonal polynomials of Hermite, Gegenbauer and Laguerre appear naturally in 
the wave functions of the principal quantum mechanics models for two reasons, except for their orthogonality. They are 
solutions of second order differential equations, so that these polynomials are eigenfunctions of self adjoint Sturm-Liouville 
differential operator. If we think about the matrix interpretation of quantum mechanics, the vectors composed by the orthogonal 
polynomials are eigenvectors of Jacobi operators. Both eigenproblems, for the Sturm-Liouville and the Jacobi operators, 
resemble the general time-independent Shr\"odingier equation. 

On the other hand, the multi-particle wave functions are either antisymmetric in the case of fermions, because of the Pauli 
exclusion principle, or symmetric when the particles are bosons. So the natural choice for the multi-particle wave functions 
are the determinants and the permanents, respectively. It is very surprising that the polynomials $\mathcal{W}_n$ appear 
so naturally in the context of the Lee-Yang measures and they look like wave functions. However, they are quite peculiar in the 
sense that they contain the Slater determinants, which are the wave function of fermions, but are symmetric like the 
permanents, which are the wave function of bosons. They are simply simmetrized Slater determinants. 

Before we proceed with the comments we show the stunning simplicity behind the relations between Jensen and Appell polynomials 
and the Slater determinants and the wave functions $\mathcal{W}_n$ which is revealed by an equivalent ``lifting" of the Appell polynomials. 
It follows from (\ref{An}) that 
$$
A_n (\mathcal{L}_\mu;z) = \int_{-\infty}^{\infty} \sum_{k=0}^n {n \choose k}\,  t^k  (-z)^{n-k} d\mu(t) = \int_{-\infty}^{\infty}  (t-z)^{n} d\mu(t). 
$$
Then the unique multivariate afine polynomial which extends it is 
$$
\mathcal{A}_n (\mathcal{L}_\mu;\mathbf{z}) = \int_{-\infty}^{\infty}  (t-z_1)(t-z_2)\cdots (t-z_n)\, d\mu(t).
$$
Similarly, for the Jensen polynomials we have 
\begin{equation}
g_n (\mathcal{L}_\mu;z) = \int_{-\infty}^{\infty}  (1-tz)^{n} d\mu(t)
\end{equation}
and 
$$
\mathcal{G}_n (\mathcal{L}_\mu;\mathbf{z}) = \int_{-\infty}^{\infty}  (1-t z_1)(1- t z_2)\cdots (1- t z_n)\, d\mu(t).
$$
Then (\ref{AnW}) and the fact that $W(p_1,\ldots,p_n;z)$ is extended uniquely to the symmetric wave function $\mathcal{W}_n(\mathbf{z})$, defined by (\ref{WW}),  yield
\begin{equation}
\label{WnI}
\mathcal{W}_n(\mathbf{z})   =   \frac{C_{n}}{\sqrt{m_0 D_{n}} }\,  \int_{-\infty}^{\infty}  (t-z_1)(t-z_2)\cdots (t-z_n)\, d\mu(t).
\end{equation}
Therefore 
\begin{equation}
\label{SnVn}
\mathcal{S}_n (p_1,\ldots,p_n; \mathbf{z}) = \frac{C_{n}}{\sqrt{m_0 D_{n}} }\,  V_n(\mathbf{z})  \int_{-\infty}^{\infty}  (t-z_1)(t-z_2)\cdots (t-z_n)\, d\mu(t).
\end{equation}

This reminds a formula of Chirtoffel about the polynomial of degree one which is orthogonal with respect to $(t-z_1)(t-z_2)\cdots (t-z_{n-1})\, d\mu(t)$, 
calculated at $z_n$ (see Theorem 2.5 in \cite{Sze75}).  This suggests that (\ref{SnVn}), and then the fundamental relation (\ref{WnI}), might be established 
employing Christoffel's formula and then the main results would follow form (\ref{WnI}). The difficulty is in the fact that the measure in the Theorem of Cristoffel changes sign and such an alternative proof is not necessary since we have given one.

Let us point our that (\ref{SnVn}) is a relation which is helpful for calculation of Slater determinants when the weight functions involve orthogonal polynomials. 
In fact, these examples explain why the Slater determinants which appear in this paper involve polynomials starting from the one of first degree. If we consider 
the wave functions of the harmonic oscillator, they are $\phi_n(x) = e^{-x^2/2} H_n(x)$, where $H_n(x)$ are properly normalised Hermite 
polynomials. The first wave function, which corresponds to the ground state of the energy, is $\phi_1(x)$, that is, the one that involves the Hermite polynomial of degree one.  
 
A general phenomena is that the evolution of a system in physics is described mathematically in terms of a formal Laplace transform, 
or a convolution with the normal distribution, where the variable is either the temperature or time but is inverse (reciprocal). Two convincing 
examples are the partition function is statistical mechanics and the general solution of  the initial value problem of the homogeneous heat equation (see (\ref{he}) below). The results in the present note show that the most natural way to approximate such functions 
is in terms of the wave functions $\mathcal{W}_n$, where the variable is the temperature or the time. Moreover, the zeros of the formal Laplace 
transform inherit the properties of the zeros $\mathcal{W}_n$. 

Moreover, the way the Slater determinants were symmetrized suggests that one may try to construct wave functions of multiparticle systems in quantum mechanics in the presence of both fermions and bosons following this hint. More precisely, one may construct the Slater determinant, composed by the wave functions of all particles and divide by the Vandermonde determinate of the variables which correspond to the bosons only.      

 \subsection{The Toda lattice and the heat equation}
 
 The Toda lattice was introduced by Toda (1967). It is model for a nonlinear one-dimensional crystal that describes the motion 
 of a chain of particles $N$ with nearest neighbour interactions. The  Hamiltonian of the Toda lattice is  
 $$
 H(\mathbf{p}, \mathbf{q}) = \sum_{k=1}^{N} \left( \frac{p_k^2(t)}{2} + e^{-(q_{k+1}(t) - q_{k}(t))} \right),
 $$
where $p_k$ is the moment of the $k$th particle and $q_k$ is its displacement from the equilibrium. With the change of variables 
of Flashka and Moser
$$
a_k = \frac{1}{2} e^{-(q_{k+1} - q_{k})/2},\ \ b_k = - \frac{1}{2} p_k,
$$ 
the equations of motion become 
$$
a_k^\prime(t) = a_k(t) ( b_k(t) - b_k(t) ), \ \ \ a_k^\prime(t) = 2 (a_k^2(t) - a_{k-1}^2(t)) 
$$
In order to write them in the Lax form, let us define the Jacobi matrix $L=(l_{i,j})$, with diagonal entries $l_{k,k}=b_k$ and off-diagonal 
ones $l_{k,k+1}=l_{k+1,k}=a_k$, as well as $B=(b_{i,j})=Skew(L)$, which has only off-diagonal entries $b_{k,k+1}=-b_{k+1,k}=a_k$.   
Then the Lax form of the equations of motion is 
$$
\frac{d}{d\, t} L = [B,L].
$$
The matrix $L$ is naturally associated with the sequence of parametric, with respect to the time variable, orthonormal polynomials which satisfy the three term recurrence relation 
$$
a_n(t)\, p_{n+1} (x;t) = (x-b_n(t))\, p_{n} (x;t) -  a_{n-1}(t)\, p_{n-1} (x;t).
$$
These are in fact the characteristic polynomials of the principal minors of $L$ and they are orthogonal with respect to a measure $d\mu_t(x) = e^{tx} 
d\mu_0(x)$, where $d\mu_0$ corresponds to $t=0$. Once  the direct problem with the initial data at $t=0$ is solved and the polynomials 
$p_n(x,0)$ are obtained, one needs to solve the inverse problem. A fundamental problem is to construct $p_n(x,t)$. In order to do this, 
as it is seen from (\ref{OP}), it suffices the calculate the moments $m_k(t) = \int t^k e^{tx} d\mu_0(x)$. On the other hand, the moments are obtained by successive differentiation of 
zeroth one because $d m_k(t) /dt = m_{k+1}(t)$. Therefore, the principal task in solving the inverse problem is to determine 
$$
m_0(t) = \int e^{tx} d\mu_0(x).
$$  
In principal we may think of using the formal expansion $m_0(t) = \sum_0^\infty m_k(0) t^k/k!$.  However, an alternative approach could be 
to observe that $$m_0(t) = \mathcal{L}_{\mu_0}(-t)$$ and approximate the latter formal Laplace transform by the Wronskians of the orthogonal 
polynomials $p_n(x;0)$ as in Theorem 1.

The idea of approximation by Wronskians, or more generally, by $\mathcal{W}_n$ may be applied to construct convergent 
approximants, on the basis of orthogonal polynomials, of the general 
solution 
\begin{equation}
\label{he}
u(x;t) = \frac{1}{\sqrt{4\pi k t}} \int_{-\infty}^{\infty} \exp \left( - \frac{(x-y)^2}{4kt} \right) g(y)\, dy 
\end{equation}
of the initial value problem for the  homogeneous heat equation
\begin{eqnarray*}
u_t(x;t) & = & k\, u_{xx}(x;t),\ \ \ \ (x,t) \in \mathbb{R} \times (0,\infty),\\
u(x;0) & = & g(x).
\end{eqnarray*}
 
 \subsection{The Riemann $\xi$ function and a statistical mechanics model}  A natural conjecture which has been discussed in the literature is 
 that the function $\xi(iz)$, where $\xi$ is the Riemann $\xi$ function, is a partition function of a general Ising model related to a quantum field theory 
 (see  \cite{New91}) and the references therein). We add two general comments. The first one concerns the fact that the model should be of quantum flavour. Indeed, the kernel $\Phi(t)$ is expressed, in a straightforward manner, in terms of the the Jacobi theta function $\theta_3(0|\tau)$, with exponential argument $\tau= i e^{2t}$, where 
 $$
 \theta_3(z | \tau) = \sum_{k=-\infty}^{\infty} \exp (k^2 \pi i \tau +  2 k i z).
 $$             
On the other hand, $\theta_3(z | -2 \hbar t /(\pi m))$ is a solution of the univariate time dependent Shr\"dingier equation. 

Theorem 2 provides necessary and sufficient conditions on the corresponding model, related to a quantum field theory, in terms 
of general correlation inequalities. The fact that the argument of the $\theta$-function is exponential suggests that the actions in the model 
must be exponential, too. This is justified also by the Montogomery-Dyson phenomena that (under the Riemann hypothesis) certain correlation 
functions which involve logarithms of the zeros of the $\xi$ function  behave like the same correlation function of  a random matrix from the Gaussian unitary ensemble (see \cite{Dei}).

\end{document}